\documentclass[a4paper]{jpconf}
\usepackage{graphicx}
\begin{document}

\title{Collective properties of nucleus-nucleus collisions from AGS to LHC energies}

\author{V P Konchakovski$^1$, V D Toneev$^2$, W Cassing$^1$, E L Bratkovskaya$^3$, \\ S A Voloshin$^4$ and V Voronyuk$^5$}
\address{$^1$ Institute for Theoretical Physics, University of Giessen, Giessen, Germany}
\address{$^2$ Joint Institute for Nuclear Research, Dubna, Russia}
\address{$^3$ Institute for Theoretical Physics, University of Frankfurt, Frankfurt, Germany}
\address{$^4$ Wayne State University, Detroit, Michigan, USA}
\address{$^5$ Bogolyubov Institute for Theoretical Physics, Kiev, Ukraine}
\ead{Wolfgang.Cassing@theo.physik.uni-giessen.de}

\begin{abstract}
The azimuthal anisotropies of the collective transverse flow of
charged hadrons are investigated in a wide range of heavy-ion
collision energies within the microscopic Parton-Hadron-String
Dynamics (PHSD) transport approach which incorporates explicit
partonic degrees-of-freedom in terms of strongly interacting
quasiparticles (quarks and gluons) in line with an equation-of-state
from lattice QCD as well as the dynamical hadronization and hadronic
collision dynamics in the final reaction phase. The experimentally
observed increase of the elliptic flow $v_2$ of charged hadrons with
collision energy is successfully described in terms of the PHSD
approach.  The analysis of higher-order harmonics $v_3$ and $v_4$ in
the azimuthal angular distribution shows a similar tendency of
growing deviations between partonic and purely hadronic models with
increasing collision energy. This demonstrates that the excitation
functions of azimuthal anisotropies reflect the increasing role of
quark-gluon degrees of freedom in the early phase of relativistic
heavy-ion collisions. Furthermore, the specific variation of the
ratio $v_4/(v_2)^2$ with respect to bombarding energy, centrality
and transverse momentum is found to provide valuable information on
the underlying partonic dynamics.
\end{abstract}

\section{Introduction}
The discovery of large azimuthal anisotropic flow at the
Relativistic-Heavy-Ion-Collider (RHIC) provides a conclusive evidence
for the creation of dense partonic matter in ultra-relativistic
nucleus-nucleus collisions. With sufficiently strong parton
interactions, the medium in the collision zone can be expected to
achieve local equili\-brium and exhibit approximately hydrodynamic
flow~\cite{Ol92,HK02,Sh09}. The momentum anisotropy is generated due
to pressure gradients of the initial ``almond-shaped'' collision zone
produced in noncentral collisions~\cite{Ol92,HK02}.  The azimuthal
pressure gradient extinguishes itself soon after the start of the
hydrodynamic evolution, so the final flow is only weakly sensitive to
later stages of the fireball evolution. The pressure gradients have to
be large enough to translate an early asymmetry in density of the
initial state to a final-state momentum-space anisotropy. In these
collisions a new state of strongly interacting matter is created,
being characterized by a very low shear viscosity $\eta$ to entropy
density $s$ ratio, $\eta/s$, close to a nearly perfect
fluid~\cite{Sh05,GMcL05,PC05}.  Lattice QCD (lQCD)
calculations~\cite{Aoki:2006we,Aoki:2006br,Borsanyi:2010bp} indicate
that a crossover region between hadron and quark-gluon matter should
have been reached in these experiments.

An experimental manifestation of this collective flow is the
anisotropic emission of charged particles in the plane transverse to
the beam direction. This anisotropy is described by the different
flow parameters defined as the proper Fourier coefficients $v_n$ of
the particle distributions in azimuthal angle $\psi$ with respect to
the reaction plane angle $\Psi_{RP}$. At the highest RHIC collision
energy of $\sqrt{s_{NN}} =$ 200~GeV, differential elliptic flow
measurements $v_2(p_T)$ have been reported for a broad range of
centralities or number of participants $N_{part}$. For $N_{part}$
estimates, the geometric fluctuations associated with the positions
of the nucleons in the collision zone serve as the underlying origin
of the initial eccentricity fluctuations. These data are found to be
in accord with model calculations that an essentially locally
equilibrated quark gluon plasma (QGP) has little or no
viscosity~\cite{PHEN07,RR07,XGS08}. Collective flow continues
to play a central role in characterizing the transport properties of
the strongly interacting matter produced in heavy-ion collisions at
RHIC. Particle anisotropy measurements are considered as key
observables for a reliable extraction of transport coefficients.

It was shown before that higher-order anisotropy harmonics, in
particular $v_4$, may provide a more sensitive constraint on the
magnitude of $\eta/s$ and the freeze-out dynamics, and the ratio
$v_4/(v_2)^2$ might indicate whether a full local equilibrium is
achieved in the QGP~\cite{Bh05}. The role of fluctuations and
so-called `nonflow' correlations are important for such
measurements. It is well established that initial eccentricity
fluctuations significantly influence the magnitudes of
$v_{2,4}$~\cite{MS03,MS03a}. However, the precise role of nonflow
correlations, which lead to a systematic error in the determination
of $v_{2,4}$, is less clear. Recently, significant attention has
been given to the study of the influence of initial geometry
fluctuations on higher order eccentricities $\epsilon_n (n\geq 3$)
for a better understanding of how such fluctuations manifest
themselves in the harmonic flow correlations characterized by $v_n$.
Even more, it was proposed that the analysis of $v_n^2$ for all
values of $n$ can be considered as an analogous measurement to the
Power Spectrum extracted from the Cosmic Microwave Background
Radiation providing a possibility to observe superhorizon
fluctuations~\cite{MMSS08}.

A large number of anisotropic flow measurements have been performed
by many experimental groups at SIS, AGS, SPS and RHIC energies over
the last twenty years. Very recently, the azimuthal asymmetry has
been measured also at the LHC~\cite{LHC}. However, the fact that
these data have not been obtained under the same experimental
conditions as at RHIC experiments, does not directly allow for a
detailed and meaningful comparison in most cases. The experimental
differences include: different centrality selection, different
transverse momentum acceptance, different particle species,
different rapidity coverage and different methods for flow analysis
as pointed out in Ref.~\cite{Tar11}.

The Beam-Energy-Scan (BES) program proposed at RHIC~\cite{ST11}
covers the energy interval from $\sqrt{s_{NN}}=$ 200~GeV, where
partonic degrees of freedom play a decisive role, down to the AGS
energy of $\sqrt{s_{NN}}\approx$ 5~GeV, where most experimental data
may be described successfully in terms of hadronic
degrees-of-freedom, only. Lowering the RHIC collision energy and
studying the energy dependence of anisotropic flow allows to search
for the possible onset of the transition to a phase with partonic
degrees-of-freedom at an early stage of the collision as well as
possibly to identify the location of the critical end-point that
terminates the cross-over transition at small quark-chemical
potential to a first order phase transition at higher quark-chemical
potential~\cite{Lac07,Agg07}.

This contribution aims to summarize excitation functions for
different harmonics of the charged particle anisotropy in the
azimuthal angle at midrapidity in a wide transient energy range,
{\it i.e.} from the AGS to the top RHIC energy. The first attempts
to explain the preliminary STAR data with respect to the observed
increase of the elliptic flow $v_2$ with the collision energy have
failed since the traditional available models did not allow to
clarify the role of the partonic phase~\cite{NKKNM10}. In this
contribution  we investigate the energy behavior of different flow
coefficients, their scaling properties and differential
distributions (cf.  Ref.~\cite{v2short,v2long}). Our analysis of the
STAR/PHENIX RHIC data -- based on recent results of the BES program
-- will be performed within the Parton-Hadron-String Dynamics (PHSD)
transport model~\cite{PHSD} that includes explicit partonic
degrees-of-freedom as well as a dynamical hadronization scheme for
the transition from partonic to hadronic degrees-of-freedom and vice
versa. For more detailed descriptions of PHSD and its ingredients we
refer the reader to Refs. \cite{BCKL11,Cassing06,Cassing07,Cas09}.

\section{Results for collective flows}

We directly continue with the results from PHSD in comparison with
other approaches and the available experimental data.

\subsection{Elliptic flow}

The largest component, known as elliptic flow $v_2$, is one of the
early observations at RHIC~\cite{Ac01}.
 The elliptic flow coefficient is a widely used quantity characterizing
the azimuthal anisotropy of emitted particles,
 \begin{equation} \label{eqv2}
 v_2 = <cos(2\psi-2\Psi)>=<\frac{p^2_x - p^2_y}{p^2_x + p^2_y}>~,
\end{equation}
where $\Psi_{RP}$ is the azimuth of the reaction plane, $p_x$ and
$p_y$ are the $x$ and $y$ component of the particle momenta and the
brackets denote averaging over particles and events. This
coefficient can be considered as a function of centrality,
pseudorapidity $\eta$ and/or transverse momentum $p_T$. We note that
the reaction plane in PHSD is given by the $(x-z)$ plane with the
$z$-axis in beam direction. The reaction plane is defined as a plane
containing the beam axes and the impact parameter vector.

\begin{figure}[t]
\includegraphics[width=8cm,height=6cm]{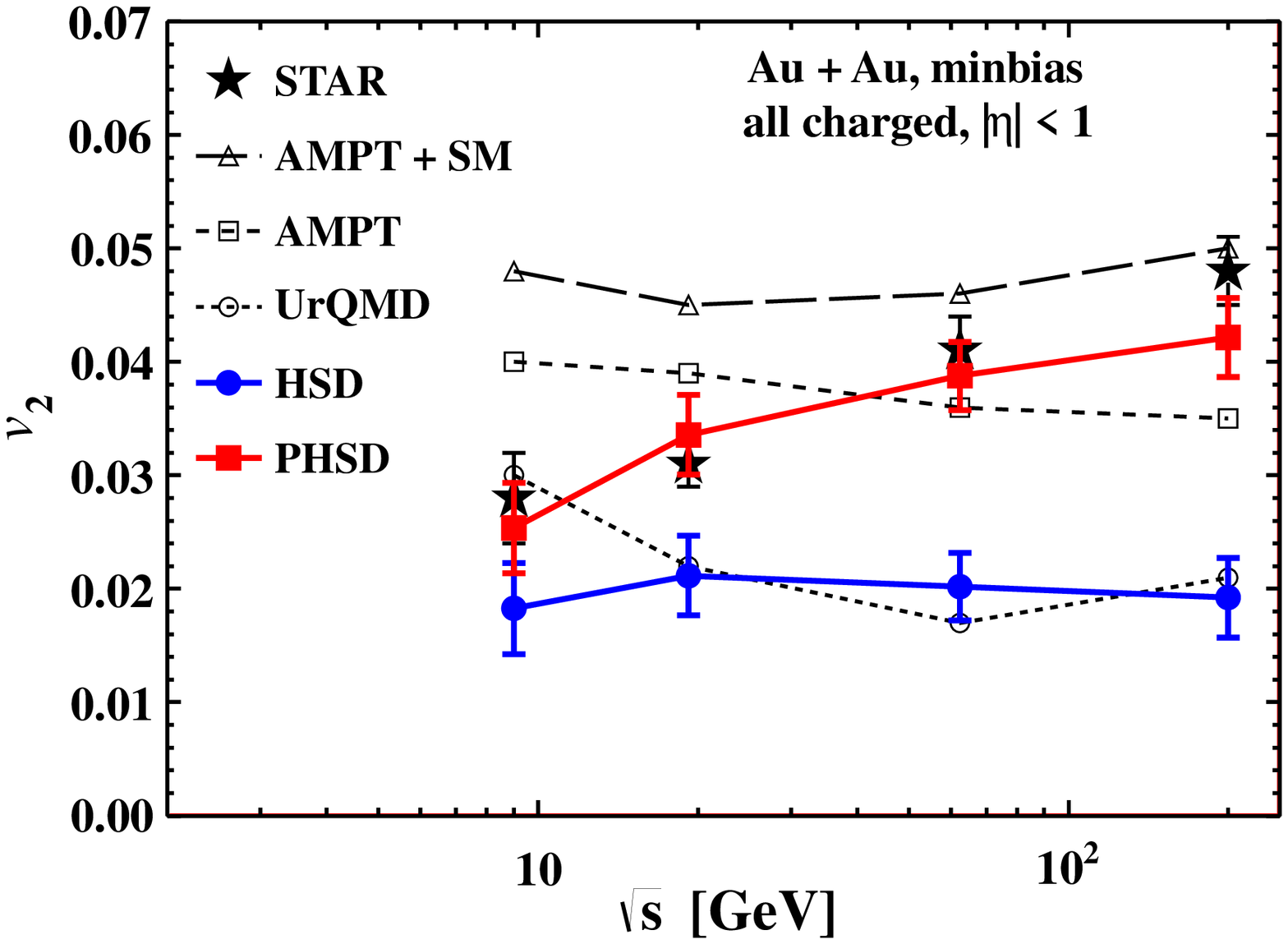}\includegraphics[width=8cm,height=6cm]{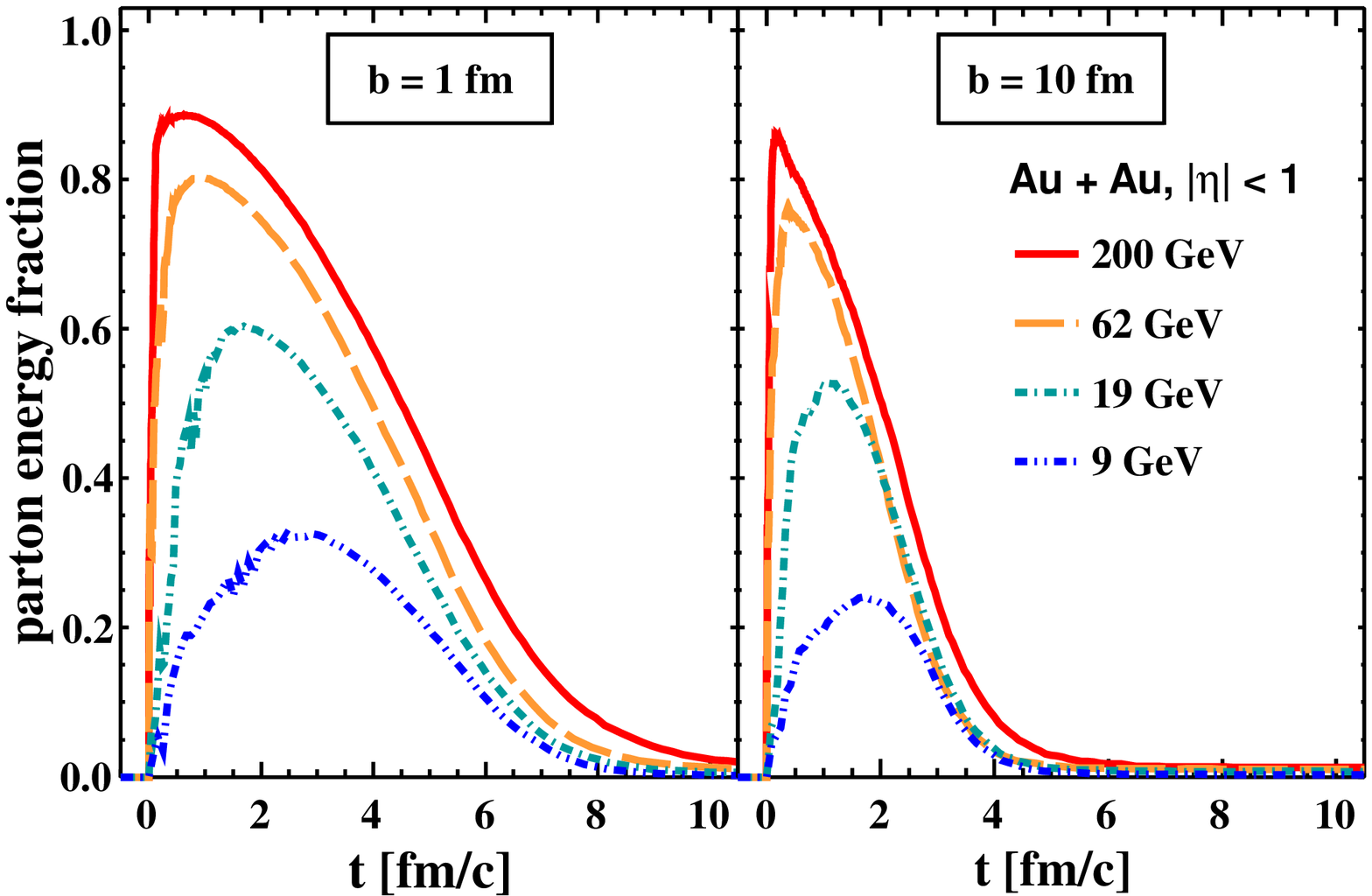}
 \caption{(l.h.s.) The average elliptic flow $v_2$ of
charged particles at
  midrapidity for minimum bias collisions at $\sqrt{s_{NN}}=$ 9.2,
  19.6, 62.4 and 200~GeV (stars) is taken from the data
  compilation of Ref.~\cite{NKKNM10}). The corresponding results from
  different models are compared to the data and explained in more
  detail in the text.  (r.h.s.) Evolution of the parton fraction of the
total energy density
  at midrapidity (from PHSD) for different collision energies at impact parameters
  $b=$1 fm and 10 fm.}
\label{s}
\end{figure}

We recall that at high bombarding energies the longitudinal size of
the Lorentz contracted nuclei becomes negligible compared to its
transverse size. The forward shadowing effect then becomes negligible and the
elliptic flow fully develops in-plane, leading to a positive value
of the average flow $v_2$ since no shadowing from spectators takes
place. In Fig.~\ref{s} (l.h.s.) the experimental $v_2$ data compilation for
the transient energy range is compared to the results from HSD
calculations and further available model results as included in
Ref.~\cite{NKKNM10}. The centrality selection is the same for the
data and the various models.

In order to interpret the results in Fig.~\ref{s} we have to recall
the various ingredients of the models employed for comparison. The
UrQMD (Ultra relativistic Quantum Molecular Dynamics) model is a
microscopic transport theory based on the relativistic Boltzmann
equation~\cite{PR90,UrQMD}. It allows for the on-shell propagation of all
hadrons along classical trajectories in combination with stochastic
binary scattering, color string formation and resonance decay. The
model incorporates baryon-baryon, meson-baryon and meson-meson
interactions based on experimental data (when possible). This
Boltzmann-like hadronic transport model has been employed for
proton-nucleus and nucleus-nucleus collisions from AGS to RHIC
energies~\cite{UrQMD}. The comparison of the data on $v_2$ to those
from the UrQMD model will thus essentially provide information on
the contribution from the hadronic phase. As seen in Fig.~\ref{s},
being in agreement with data at the lowest energy $\sqrt{s_{NN}}=$
9.2~GeV, the UrQMD model results then either remain approximately
constant or decrease slightly with increasing $\sqrt{s_{NN}}$; UrQMD
thus does not reproduce the rise of $v_2$ with the collision energy
as seen experimentally.

The HSD model~\cite{BCS03,Ehehalt,HSD} is also a hadron-string model
including formally the same processes as UrQMD. However, being based
on the off-shell generalized transport equation~\cite{Cas09}
followed from Kadanoff-Baym approach, the quasiparticles in the HSD
model take into account in-medium modifications of their properties
in the nuclear environment which is rather essential for many
observables and in particular for dileptons. Detailed comparisons
between HSD and UrQMD for central Au+Au (Pb+Pb) collisions have been
reported in Ref.~\cite{BRAT04} from AGS to top SPS energies with
respect to a large experimental data set. Indeed, both hadronic
approaches yield similar results on the level of 20-30\% which is
also the maximum deviation from the data sets. Accordingly, the HSD
model also predicts an approximately energy-independent flow $v_2$
in quite close agreement with the UrQMD results. We may thus
conclude that the rise of $v_2$ with bombarding energy is not do to
hadronic interactions and models with partonic degrees-of-freedom
have to be addressed.

The AMPT (A Multi Phase Transport model)~\cite{AMPT,AMPT2} uses
initial conditions of a perturbative QCD (pQCD) inspired model which
produces multiple minijet partons according to the number of binary
initial nucleon-nucleon collisions. These (massless) minijet partons
undergo scattering (without potentials) before they are allowed to
fragment into hadrons. The string melting (SM) version of the AMPT
model (labeled in Fig.~\ref{s} as AMPT-SM) is based on the idea that
the existence of strings (or hadrons) is impossible for energy
densities beyond a critical value of $\varepsilon\sim$ 1~GeV/fm$^3$.
Hence they  melt the strings to (massless) partons. This is done by
converting the mesons to a quark and anti-quark pair, baryons to
three quarks {\it etc}. fulfilling energy-momentum conservation. The
subsequent scattering of the quarks are based on a parton cascade
with (adjustable) effective cross sections which are significantly
larger than those from pQCD~\cite{AMPT,AMPT2}.  Once the partonic
interactions terminate, the partons hadronize through the mechanism
of parton coalescence.

We find from Fig.~\ref{s} that the interactions between the minijet
partons in the AMPT model indeed increase the elliptic flow
significantly as compared to the hadronic models UrQMD and HSD. An
additional inclusion of interactions between partons in the AMPT-SM
model gives rise to another 20\% of $v_2$ bringing it into agreement
(for AMPT-SM) with the data at the maximal collision energy. So,
both versions of the AMPT model indicate the importance of partonic
contributions to the observed elliptic flow $v_2$ but do not
reproduce its growth with $\sqrt{s_{NN}}$. The authors address this
result to the partonic-equation-of state (EoS) employed which
corresponds to a massless and noninteracting relativistic gas of
particles. This EoS deviates severely from the results of lattice
QCD calculations for temperatures below 2-3 $T_c$. Accordingly, the
degrees-of-freedom are propagated without self-energies and a parton
spectral function.

The PHSD approach incorporates the latter medium effects in line
with a lQCD equation-of-state  and also includes a dynamical
hadronization scheme based on covariant transition rates. As has
been demonstrated in Refs. \cite{v2short,v2long} and explicitly
shown in Fig.~\ref{s} (l.h.s.), the elliptic flow $v_2$ from PHSD (red line)
agrees  with the data from the STAR  collaboration and clearly shows
an increase with bombarding energy. Note that PHSD and AMPT-SM
practically give the same elliptic flow at the top RHIC energy of
$\sqrt{s_{NN}}=$ 200~GeV.

An explanation for the increase in $v_2$ with collision energy is
provided in Fig.~\ref{s} (r.h.s.). Here we show the partonic
fraction of the energy density with respect to the total energy
where the energy densities are calculated at mid-rapidity. As
discussed above the main contribution to the elliptic flow is coming
from an initial partonic stage at high $\sqrt{s}$. The fusion of
partons to hadrons or, inversely, the melting of hadrons to partonic
quasiparticles occurs when the local energy density is about
$\varepsilon\approx$ 0.5~GeV/fm$^3$. As follows from
Fig.~\ref{s}, the parton fraction of the total energy goes down
substantially with decreasing bombarding energy while the duration
of the partonic phase is roughly the same. The maximal fraction
reached is the same in central and peripheral collisions but the
parton evolution time is shorter in peripheral collisions. One
should recall again the important role of the repulsive mean-field
for partons in the PHSD model  that leads to
an increase of the flow $v_2$ with respect to HSD predictions (cf.
also Ref.~\cite{CB08}). We point out in addition that the increase
of $v_2$ in PHSD relative to HSD is also partly due to the higher
interaction rates in the partonic medium because of a lower ratio of
$\eta/s$ for partonic degrees-of-freedom at energy densities above
the critical energy density than for hadronic media below the
critical energy density~\cite{Mattiello,Bass}. The relative increase
in $v_3$ and $v_4$ in PHSD essentially is due to the higher partonic
interaction rate and thus to a lower ratio $\eta/s$ in the partonic
medium which is mandatory to convert initial spacial anisotropies to
final anisotropies in momentum space~\cite{Pet4}.

\subsection{Higher-order flow harmonics}

Depending on the location of the participant nucleons in the nucleus
at the time of the collision, the actual shape of the overlap area
may vary: the orientation and eccentricity of the ellipse defined by
the participants fluctuates from event to event.  Note, however,
that by averaging over many events an almond shape is regained for
the same impact parameter.

\begin{figure}[t]
\includegraphics[width=8.2cm]{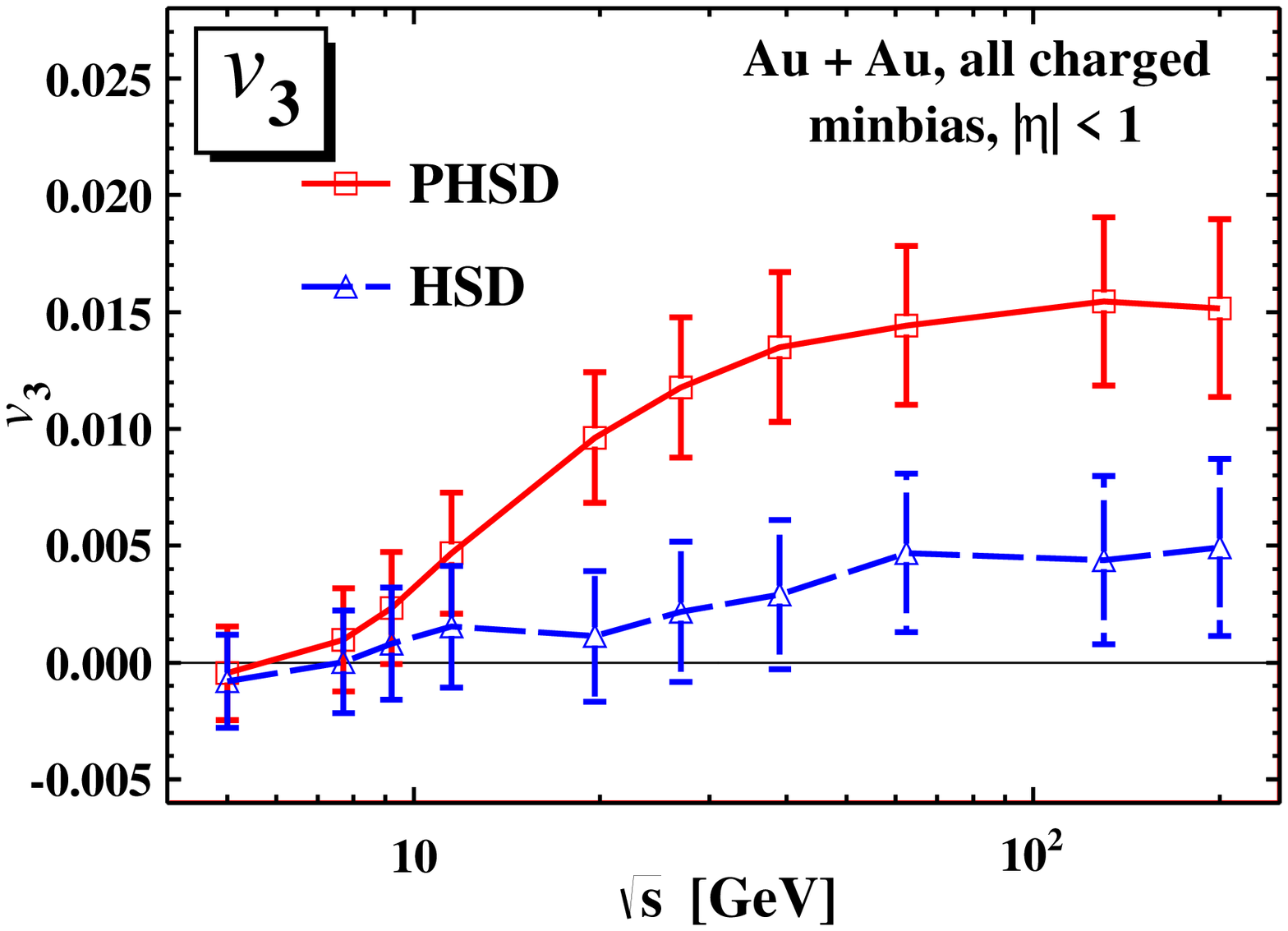}\includegraphics[width=8.2cm]{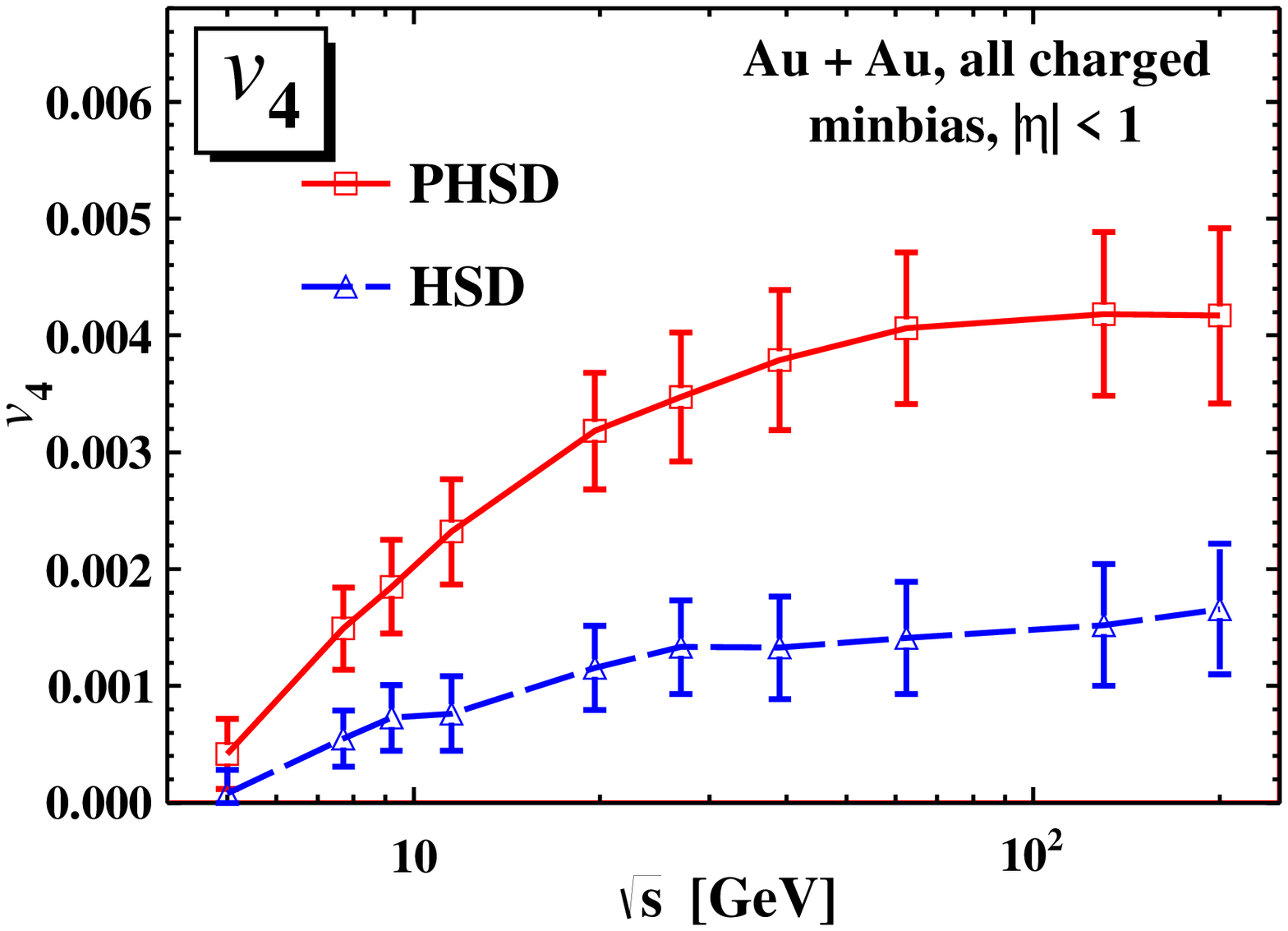}
\caption{Average anisotropic flows $v_3$ (l.h.s.) and $v_4$ (r.h.s.)
of charged
  particles at mid-pseudorapidity for minimum bias Au + Au collisions
  calculated within the PHSD (solid lines, red) and HSD (dashed lines, blue)
  models.}
\label{vns34}
\end{figure}

Recent studies suggest that fluctuations in the initial state
geometry can generate higher-order flow
components~\cite{PHEN07,MMSS08,Pet123,AR10}. The azimuthal momentum
distribution of the emitted particles is commonly expressed in the
form of Fourier series as
\begin{equation}
E\frac{d^3N}{d^3p}= \frac{d^2N}{2\pi
p_Tdp_Tdy}\left(1+\sum^\infty_{n=1} 2v_n(p_T) \cos
(n(\psi-\Psi_n))\right),\ \ \ \label{eqvn} \end{equation}
where $v_n$ is the magnitude of the $n$-th order harmonic term
relative to the angle of the initial-state spatial plane of symmetry
$\Psi_n$. The anisotropy in the azimuthal angle $\psi$ is usually
characterized by the even order Fourier coefficients with the
reaction plane $\Psi_n=\Psi_{RP}$: $v_n =\langle \exp(\, \imath \,
n(\psi-\Psi_{RP}))\rangle\ ( n = 2, 4, ...)$, since for a smooth
angular profile the odd harmonics vanish. For the odd components,
e.g. $v_3$, one should take into account event-by-event fluctuations
with respect to the participant plane $\Psi_n=\Psi_{PP}$. We
calculate the $v_3$ coefficients with respect to $\Psi_3$ as:
$v_3\{\Psi_3\} = \langle \cos(3[\psi-\Psi_3])\rangle/Res(\Psi_3)$.
The event plane angle $\Psi_3$ and its resolution $Res(\Psi_3)$ are
calculated as described in Ref.~\cite{{AdPH11}} via the
two-sub-events method~\cite{PV98,corrV2}.

In Fig.~\ref{vns34} we display the PHSD and HSD results for the
anisotropic flows $v_3$ and $v_4$ of charged particles at
mid-pseudorapidity for Au+Au collisions as a function of
$\sqrt{s_{NN}}$. The pure hadronic model HSD gives $v_3\approx$ 0
for all energies. Accordingly, the results from PHSD (dashed red
line) are systematically larger than from HSD (dashed blue line).
Unfortunately, our statistics are not good enough to allow for more
precise conclusions. The hexadecupole flow $v_4$ stays almost
constant in the energy range $\sqrt{s_{NN}}\ge$ 10~GeV; at the same
time the PHSD gives noticeably higher values than HSD which we
attribute to the higher interaction rate in the partonic phase, i.e.
a lower ratio of $\eta/s$ in the partonic
phase~\cite{Mattiello,Bass}.

\begin{figure}[t]
\includegraphics[width=8.1cm]{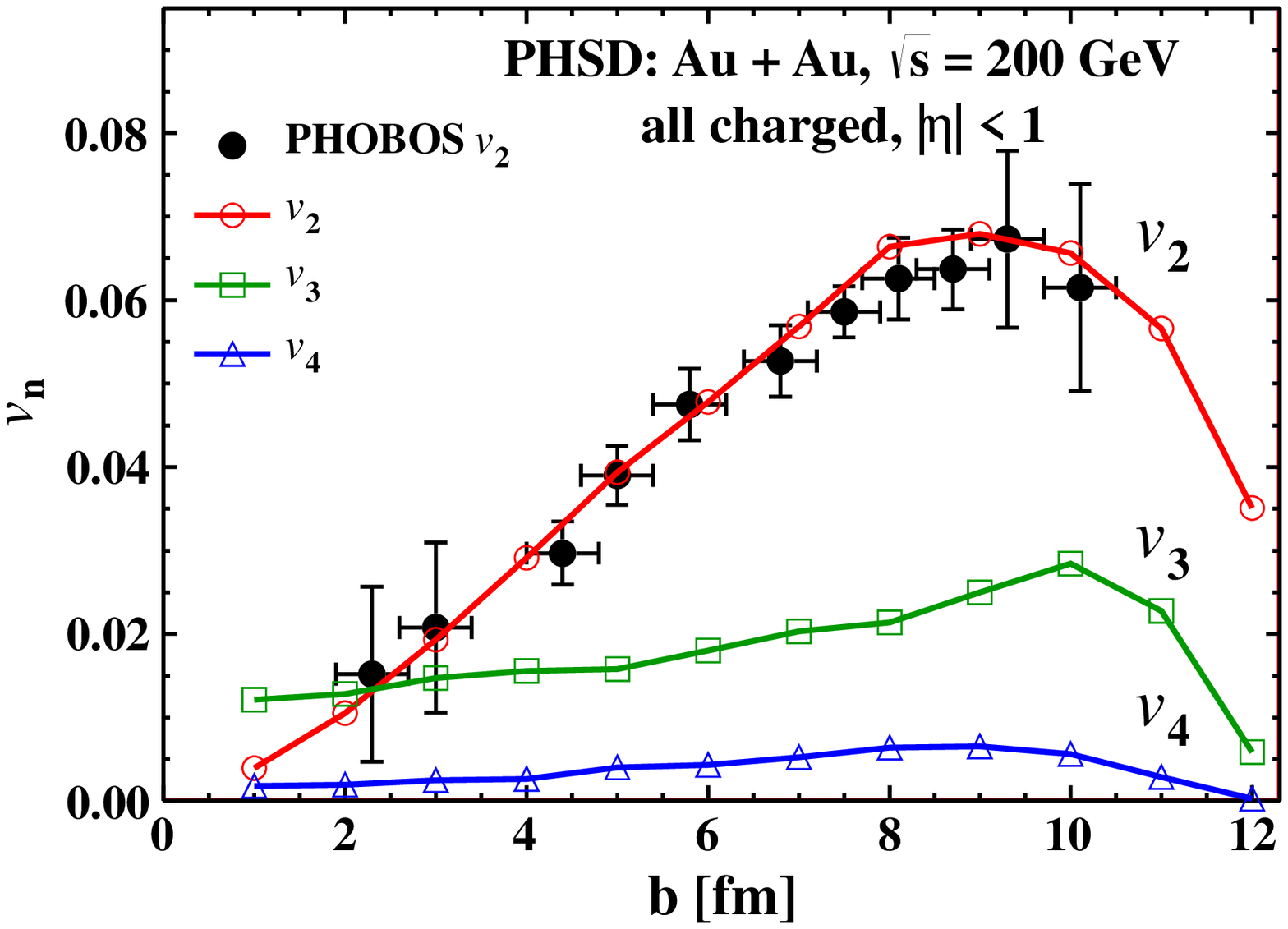}\includegraphics[width=8cm]{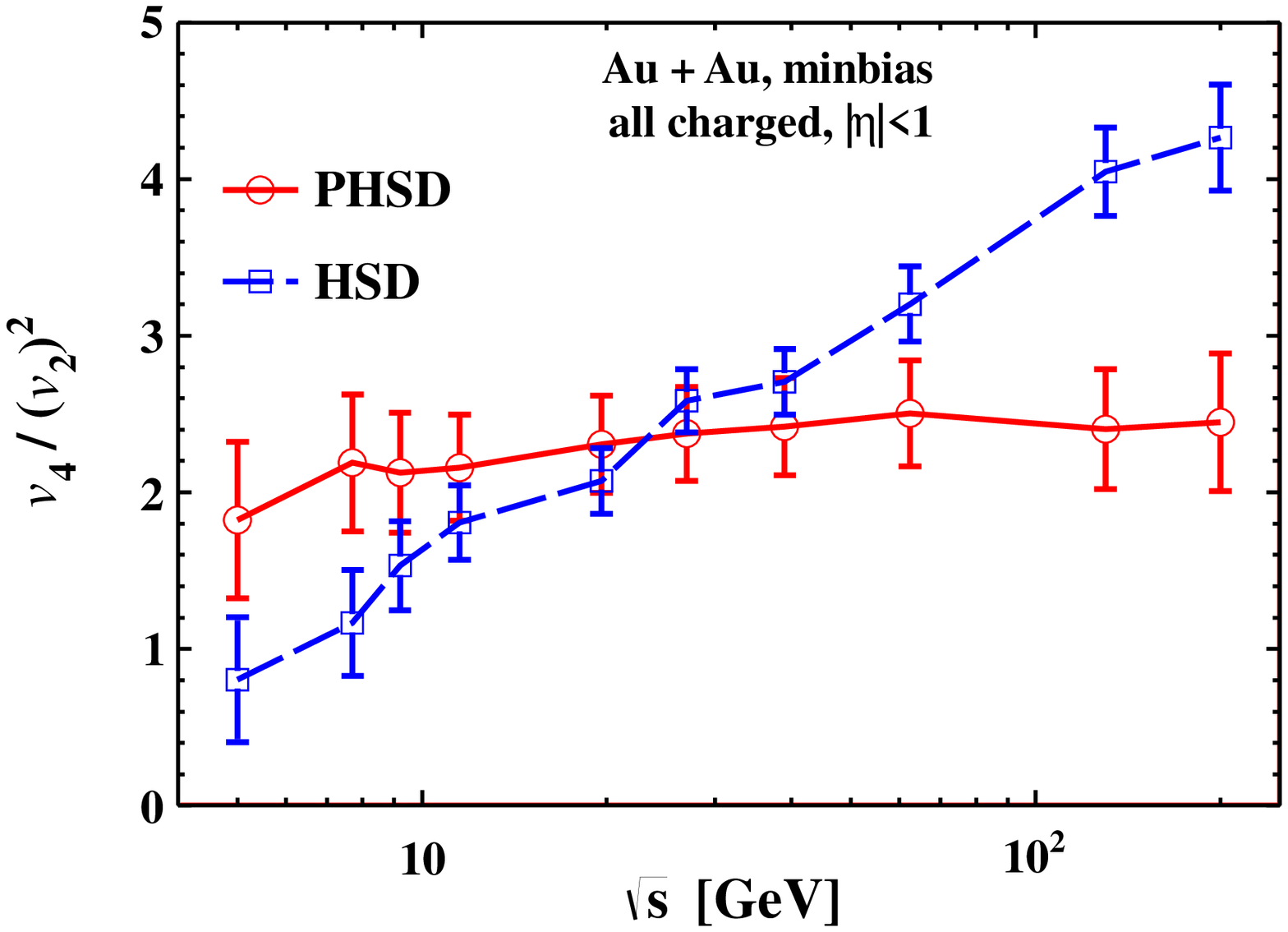}
\caption{(l.h.s.) Impact parameter dependence of anisotropic flows
of charged
  particles at mid-pseudorapidity for minimum bias collisions of Au+Au
  at $\sqrt{s_{NN}}=$ 200~GeV. Experimental points are from
  Ref.~\cite{PHO05}. (r.h.s.) Beam energy dependence of the ratio
$v_4/(v_2)^2$ for Au+Au
  collisions. The solid and dashed curves are calculated within the
  PHSD and HSD models, respectively.} \label{vnb}
\end{figure}

Alongside with the integrated flow coefficients $v_n$ the PHSD model
reasonably describes their distribution over centrality or impact
parameter $b$. A specific comparison at $\sqrt{s_{NN}}=$ 200~GeV is
shown in Fig.~\ref{vnb} for $v_2, v_3$ and $v_4$. While $v_2$
increases strongly with $b$ up to peripheral collisions, $v_{3}$ and
$v_{4}$ are only weakly sensitive to the impact parameter. The
triangular flow is always somewhat higher than the hexadecupole flow
in the whole range of impact parameters $b$.

\subsection{Ratios of different harmonics}

Different harmonics can be related to each other. In particular,
hydrodynamics predicts that $v_4 \propto (v_2)^2$~\cite{Ko03}. The
simplest prediction that $v_4 = 0.5 (v_2)^2$ is given for a boosted
thermal freeze-out distribution of an ideal fluid, Ref.~\cite{BO06}.
In this work it was noted also that $v_4$ is largely generated by an
intrinsic elliptic flow (at least at high $p_T$) rather than the
fourth order moment of the fluid flow. This is a motivation for
studying the ratio $v_4/(v_2)^2$ rather than $v_4$ alone. As is seen
in Fig. 4 (r.h.s.), indeed the ratio calculated within the PHSD
model is practically constant in the whole range of $\sqrt {s_{NN}}$
considered but significantly deviates from the ideal fluid estimate
of 0.5. In contrast, neglecting dynamical quark-gluon
degrees-of-freedom in the HSD model, we obtain a monotonous growth
of this ratio.

\begin{figure}[t]
\includegraphics[width=8.2cm]{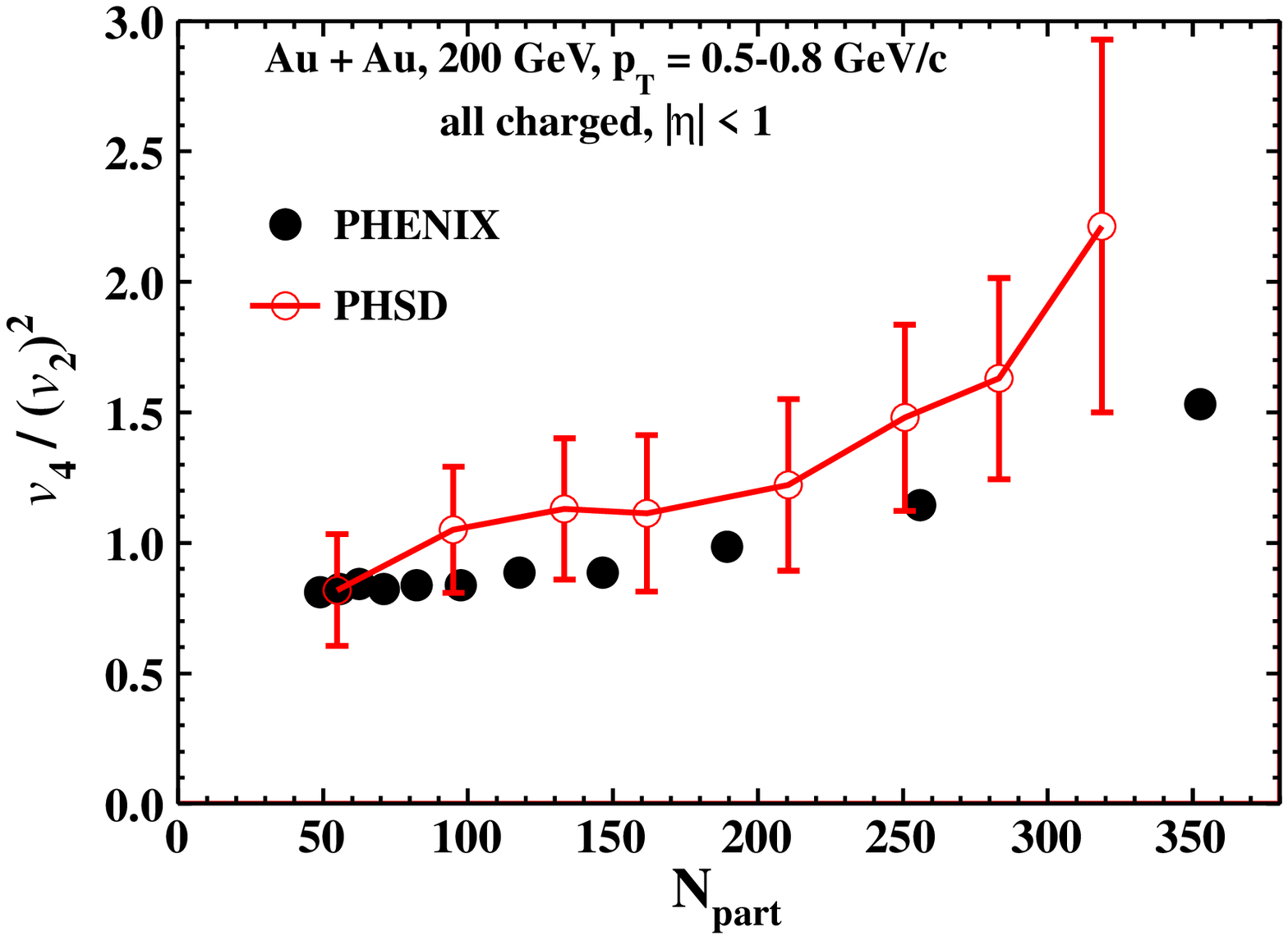}\includegraphics[width=8.2cm]{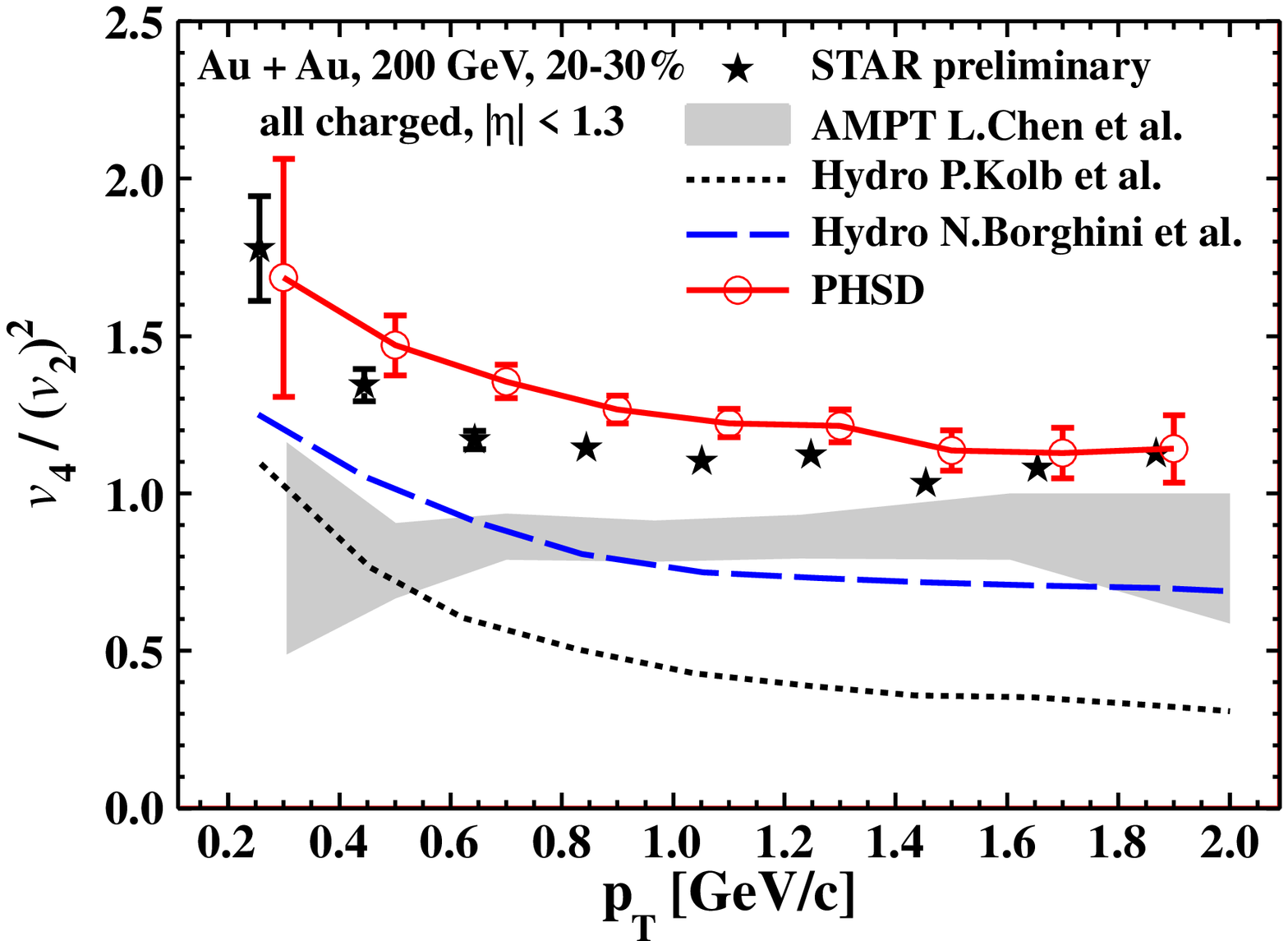}
\caption{(l.h.s.)
 Participant number dependence of the
$v_4/(v_2)^2$ ratio of
  charged particles for Au+Au ($\sqrt{s_{NN}}=$ 200~GeV) collisions.
  The experimental data points for 0.5$<p_T<$0.8~GeV/c are from
  Ref.~\cite{PHENIX-v2-s}. \\ (r.h.s.) Transverse momentum dependence of the ratio
$v_4/(v_2)^2$ of
  charged particles for Au+Au (at $\sqrt{s_{NN}}=$ 200~GeV)
  collisions. The dashed and dot-dashed lines are calculated within
  the hydrodynamic approaches from Refs.~\cite{Ko03} and~\cite{BO06},
  respectively. The shaded region corresponds to the results from the
  AMPT model~\cite{CKL04}. The experimental data points are from the
  STAR Collaboration~\cite{Bai07}.}
\label{v422Np}
\end{figure}

The dependence of the $v_4/(v_2)^2$ ratio versus the number of
participants $N_{part}$ is shown in Fig.~\ref{v422Np} for charged
particles produced in Au + Au collisions at $\sqrt{s_{NN}}=$
200~GeV. The PHSD results are roughly in agreement with the
experimental data points from Ref.~\cite{Bai07} but overshoot them
for $N_{part}\sim$ 250.

As pointed out before, the ratio $v_4/(v_2)^2$ is sensitive to the
microscopic dynamics. In this respect we show the transverse
momentum dependence of the ratio $v_4/(v_2)^2$ in Fig.~\ref{v422Np}
for charged particles produced in Au+Au collisions at
$\sqrt{s_{NN}}=$ 200~GeV (20-30\% centrality). The PHSD results are
quite close to the experimental data points from Ref.~\cite{Bai07},
however, overestimate the measurements by up to 20\%. The
hydrodynamic results -- plotted in the same figure -- significantly
underestimate the experimental data and noticeably depend on
viscosity. The partonic AMPT model~\cite{CKL04} discussed above also
predicts a slightly lower ratio than the measured one, however,
being in agreement with both hydrodynamic models for $p_T\lsim$
0.8~GeV/c. Our interpretation of Fig.~\ref{v422Np} (r.h.s.) is as
follows: the data are not compatible with ideal hydrodynamics and a
finite shear viscosity is mandatory (in viscous hydrodynamics) to
come closer the experimental observations. The kinetic approaches
AMPT and PHSD perform better but either overestimate (in AMPT) or
slightly underestimate the scattering rate of soft particles (in
PHSD). An explicit study of the centrality dependence of these
ratios should provide further valuable information.

\section{Conclusions}

In summary, relativistic collisions of Au+Au from $\sqrt{s_{NN}}=$ 5
to 200~GeV have been studied within the PHSD approach which includes
the dynamics of explicit partonic degrees-of-freedom as well as
dynamical local transition rates from partons to hadrons and also
the final hadronic scatterings. Whereas earlier studies have been
carried out for longitudinal rapidity distributions of various
hadrons, their transverse mass spectra and the elliptic flow $v_2$
as compared to available data at SPS and RHIC
energies~\cite{PHSD,BCKL11}, here we have focussed on the PHSD
results for the collective flow coefficients $ v_2, v_3$ and
$v_4$ in comparison to recent experimental data in the large energy
range from the RHIC Beam-Energy-Scan (BES) program as well as
different theoretical approaches ranging from hadronic transport
models to ideal and viscous hydrodynamics. We mention explicitly
that the PHSD model from Ref.~\cite{BCKL11} has been used for all
calculations performed in this study and no tuning (or change) of
model parameters has been performed.

We have found that the anisotropic flows -- elliptic $v_2$,
triangular $v_3$, hexadecapole $v_4$ -- are reasonably described
within the PHSD model in the whole transient energy range naturally
connecting the hadronic processes at lower energies with
ultrarelativistic collisions where the quark-gluon degrees of
freedom become dominant. The smooth growth of the elliptic flow
$v_2$ with the collision energy demonstrates the increasing
importance of partonic degrees of freedom. This feature is
reproduced by neither hadron-string based kinetic models nor A Multi
Phase Transport (AMPT) model treating the partonic phase in a
simplified manner. Other signatures of the transverse collective
flow, the higher-order harmonics of the transverse anisotropy $v_3$
and $v_4$ change only weakly from $\sqrt{s_{NN}}\sim$ 7~GeV to the
top RHIC energy of $\sqrt{s_{NN}}=$ 200~GeV, roughly in agreement
with experiment. As shown in this study, this success is related to
a consistent treatment of the interacting partonic phase in PHSD
whose fraction increases with the collision energy.

The analysis of correlations between particles emitted in
ultrarelativistic heavy-ion collisions at large relative rapidity
has revealed an azimuthal structure that can be interpreted as
solely due to collective flow~\cite{
XK11,SJG11,TY10,LGO10}. This
interesting new phenomenon, denoted as triangular flow, results from
initial state fluctuations and a subsequent hydrodynamic-like
evolution. Unlike the usual directed flow, this phenomenon has no
correlation with the reaction plane and should depend weakly on
rapidity. Event-by-event hydrodynamics~\cite{GGH11} has been a
natural framework for studying this triangular collective flow but
it has been of interest also to investigate these correlations in
terms of the PHSD model. We have found the third harmonics to
increase steadily in PHSD with bombarding energy. The coefficient
$v_3$ is compatible with zero for $\sqrt{s_{NN}} >$ 20~GeV in case
of the hadronic transport model HSD which does not develop
`ridge-like' correlations. In this energy range PHSD gives a
positive $v_3$ due to dominant partonic interactions.

Different harmonics can be related to each other and in particular,
hydrodynamics predicts that $v_4 \propto (v_2)^2$~\cite{Ko03}. In
this work it was noted also that $v_4$ is largely generated by an
intrinsic elliptic flow (at least at high $p_T$) rather than the
fourth order moment of the fluid flow. Indeed, the ratio
$v_4/(v_2)^2$ calculated within the PHSD model is approximately
constant in the whole considered range of $\sqrt {s_{NN}}$ but
significantly deviates from the ideal fluid estimate of 0.5. In
contrast, neglecting dynamical quark-gluon degrees-of-freedom in the
HSD model, we obtain a monotonous growth of this ratio.

The transverse momentum dependence of the ratio $v_4/(v_2)^2$ at the
top RHIC energy has given further interesting information ({\it cf.}
Fig. 4) by comparing the various model results to the data from
STAR which are interpreted as follows: the STAR data are not
compatible with ideal hydrodynamics and a finite shear viscosity is
mandatory (in viscous hydrodynamics) to come closer the experimental
 ratio observed. The kinetic approaches AMPT and PHSD perform better but
either overestimate (in AMPT) or slightly underestimate the
scattering rate of soft particles (in PHSD). Our findings at LHC energies are close
to those at top RHIC energy.

\end{document}